# Superconductivity up to 17 K in the high-pressure rhombohedral-I phase of ReO$_3$: a potential oxide analogy of hydride superconductors

Pengfei Shan,[1,2,6] Tenglong Lu,[1,2,6] Ziyi Liu,[1,2,6] Yuanyuan Jiao,[1,3] Pengtao Yang,[1,2] Jun Hou,[1,2] Liang Ma,[1,4] Yoshiya Uwatoko,[5] Xiaoli Dong,[1,2] Bosen Wang,[1,2] Miao Liu,[1,2],* Jianping Sun,[1,2],* and Jinguang Cheng[1,2],*

[1]Beijing National Laboratory for Condensed Matter Physics and Institute of Physics, Chinese Academy of Sciences, Beijing 100190, China
[2]School of Physical Sciences, University of Chinese Academy of Sciences, Beijing 100190, China
[3]Faculty of Science, Wuhan University of Science and Technology, Wuhan, Hubei 430065, China
[4]Key Laboratory of Materials Physics, Ministry of Education, School of Physics and Microelectronics, Zhengzhou University, Zhengzhou 450052, China
[5]Institute for Solid State Physics, University of Tokyo, 5-1-5 Kashiwanoha, Kashiwa, Chiba 277-8581, Japan
[6]These authors contributed equally.

*Correspondence: mliu@iphy.ac.cn (M. L.); jpsun@iphy.ac.cn (J. S.); jgcheng@iphy.ac.cn (J. C.)

## Abstract

As an *A*-site-vacant perovskite-type oxide, ReO$_3$ undergoes sequential pressure-driven structural transitions associated with the rotation of ReO$_6$ octahedra. The rhombohedral-I phase in the pressure range of 12-39 GPa is featured by a lattice of nearly close-packed oxygen layers intercalated with Re cations, in reminiscent of the recently discovered superhydride superconductors. A combined study of first-principles calculations and transport measurements under high pressures enabled us to discover superconductivity in the rhombohedral-I phase, and it shows a dome-shaped $T_c(P)$ with a maximum $T_c$ of 17 K at about 30 GPa. In addition to the enhanced density of states at Fermi level compared to that of the ambient phase, the vibrations of hexagonal-close-packed oxygen lattice significantly strengthen the electron-phonon coupling, which is responsible for observed superconductivity with a relatively high $T_c$. The present work thus establishes a rare case among oxide superconductors that the light-element oxygen lattice plays a crucial role in inducing superconductivity.

# INTRODUCTION

In the recently discovered superhydride superconductors such as $LaH_{10}$ [1, 2] and $SH_3$ [3], the light element, hydrogen, plays a crucial role within the framework of electron-phonon coupling (EPC) mechanism [4]. In these superconducting hydrides, hydrogen forms weak covalent network structures with hydrogen [5, 6] or other elements [7]. Such structural features impart two key characteristics for high superconducting transition temperature ($T_c$) [8]: (a) the hydrogen-derived electrons make a significant contribution to the density of states near Fermi surface ($N(E_F)$), and (b) the hydrogen-related phonon vibration modes strongly enhance EPC. Similar features were observed in other light-element superconductors, such as the $MgB_2$ with honeycomb boron-layered structure [9, 10] and the graphite intercalation compounds with carbon-layered structure [11]. However, it is a challenge for the oxides to form weak covalent network structures due to the strong electronegativity of oxygen, making it difficult to generalize these above characteristics to oxide materials. Nevertheless, metallic oxides with unique bonding behaviors, particularly having a high oxygen molar ratio, are promising candidates for achieving oxide analog of hydride superconductors.

With a high oxygen molar ratio, $ReO_3$ is a purple-colored metallic oxide with high conductivity comparable to that of copper and silver [12]. It adopts an undistorted A-site-vacant perovskite structure with corner-shared $ReO_6$ octahedra [13]. Recently, the spectroscopy measurements revealed that excellent metallic conductivity originates from the covalent mixing between Re-$t_{2g}$ and O-$p_\pi$ orbitals through the linear Re-O-Re bonds [12, 14], which raises the orbitals of oxygen to the Fermi level. ReO$_3$ has been predicted as a superconductor at ambient pressure (AP) [15], but no superconductivity was observed down to 20 mK. Intriguingly, ReO$_3$ undergoes a sequential pressure-driven structural phase transitions [16-20], including cubic-I ($Pm\bar{3}m$) (C-I) to cubic-II ($Im\bar{3}$) (C-II) at ~0.5 GPa, monoclinic ($C2/c$) at ~3 GPa, rhombohedral-I ($R\bar{3}c$) (R-I) at ~12 GPa, and rhombohedral-II ($R\bar{3}c$) (R-II) at ~39 GPa. Although extensive studies have been performed to explore high-pressure phases and the critical transition pressures, no consensus has been reached, as summarized in Table S1 of the Supplemental Information (SI). Nevertheless, the presence of the R-I phase between 12 and 39 GPa is well established.

The evolution of the crystal structure for ReO$_3$ upon compression shown in Figure. 1 is particularly intriguing, especially from the perspective of the oxygen lattice. The R-I phase emerges from the C-I phase via coupled rotations of ReO$_6$ octahedra along the ⟨111⟩ direction [17]. As seen in the lower panel of Figure. 1, the oxygen atoms in the (111) plane gradually transform from a perfect kagome lattice in the C-I phase to a distorted triangular lattice in the R-I phase, which becomes more uniform upon compression and achieves a hexagonal close-packed arrangement at a rotation angle of 30° above 30 GPa. Further compression cannot be achieved only by rotating the ReO$_6$ octahedra, and the volume reduction results in another structure transition to the R-II phase above 39 GPa [17, 18].

Such close-packed oxygen layers found in the R-I phase are not realized in the cubic phases of ReO$_3$ or in most other oxide materials, making it a rather rare case that can be described as nearly close-packed oxygen layers intercalated with small Re cations, Figure. 1. Motivated by the recent discovery of near-room-temperature superconductivity in superhydrides [1, 2], we investigate how the light-element oxygen lattice would affect the phonon spectrum and the electrical transport properties of ReO$_3$ by modifying its EPC. Here, our combined crystal structure characterization, first-principles calculations, and transport measurements under high pressures enabled us to discover superconductivity in the R-I phase with an optimal $T_c$ of 17 K near 30 GPa, where the close-packed oxygen stacking forms. Theoretical analyses, supported by experimental results, indicate that the R-I phase of ReO$_3$ can serve as a potential oxide analog of hydride superconductors.

**RESULTS**

**Structural characterizations.** First, we measured high-pressure synchrotron X-ray diffraction (SXRD) and confirmed pressure-induced sequential structural transitions of ReO$_3$. As shown in Figure. S1A, the SXRD patterns can be indexed by using the C-I, C-II, mixed C-II/R-I, and R-I phase in different pressure ranges up to 30 GPa, respectively. Some new diffraction peaks emerge at 36.1 GPa, Figure. S1B, in accordance with the occurrence of structure transition to the R-II phase mentioned above. Here, we refined the structure parameters of single-phase regions for C-II and R-I, and presented the obtained pressure-volume (*P-V*) relations in Figure. 2A. As shown in Figure. 2B, the coupled rotations of ReO$_6$ octahedra within the *ab* plane exhibit a significant anisotropic compression, i.e., the *a*-axis shrinks by more than 6% while *c*-axis is only reduced by 6‰. By assuming rigid rotations of the octahedra, the rotation angle can be estimated based on the change of the *a/c* ratio. As illustrated in Figure. 3C, the rotation angle increases gradually from 25° to 30° at 30 GPa in the R-I phase, where the close-packed oxygen stacking forms. Above 30 GPa, further compression induces the octahedra rotating inward by nearly 2°, which destabilizes the R-I phase at ~ 39 GPa and leads to the structure phase transition to the R-II phase. The above results provide a quantitative description of octahedral rotation behaviors involving the structural transitions associated with the R-I phase.

**Theoretical calculations.** The structural and electronic properties of ReO$_3$ were calculated by employing the density functional theory (DFT) method at the GGA-PBE level at selected pressures. The crystal structure was fully relaxed from 0 to 45 GPa to mimic the structural phase evolutions of ReO$_3$ under external pressures. Our DFT calculations demonstrate that the rotation of ReO$_6$ octahedra in the C-I phase along the <111> axis densifies the oxygen layers and transforms it into the R-I phase above 10 GPa, consistent with the experimental results. The ideal hexagonal close packing is realized at the rotation angle of 30° around 30 GPa. Figure. 3A-C show the calculated electron density of state (DOS) at three representative pressures, *i.e.*, 0 GPa for the cubic-I phase, 15 and 30 GPa for the R-I phase with different rotation angles. For all pressures, the Re-5*d* and O-2*p* orbitals are strongly hybridized and there is a considerable contribution of O-2*p* orbitals to the $N(E_\text{F})$. With increasing pressure gradually, the $N(E_F)$ shows a slight increase at 15 GPa and is almost doubled at 30 GPa. Thus, the R-I phase is theoretically predicted to possess enhanced $N(E_F)$ with strong hybridizations between Re-5$d$ and O-2$p$ orbitals.

In addition to $N(E_F)$, the contributions of oxygen lattice to the phonon spectrum and electron-phonon coupling (EPC) were also significantly enhanced with increasing pressure in the R-I phase. Figure. 3D-F display the calculated phonon DOS (PhDOS), Eliashberg spectral function $\alpha^2F(\omega)$ and cumulative frequency-dependent EPC constant $\lambda(\omega)$ at three representative pressures. From the PhDOS, we can see that the lower frequency phonons (< 250 cm$^{-1}$) are attributed to both Re and O, while the higher frequency phonons (> 250 cm$^{-1}$) are dominated by the vibration modes of O. In addition, a gap opens around 250 cm$^{-1}$ in the phonon spectrum of R-I phase at 30 GPa. The most prominent features revealed in Figure. 3D-F are the significantly enhanced $\alpha^2F(\omega)$ and $\lambda(\omega)$ with increasing pressure in the R-I phase. In particular, $\lambda(\omega)$ at 30 GPa in the frequency range ~250-500 cm$^{-1}$ experiences an obvious upraise, which is primarily attributed to the phonon modes of the oxygen lattice mentioned above. The phonon dispersion curves at 30 GPa were presented in Figure. S2. The phonon linewidth calculations indicate that $\lambda$ predominantly originates from the third, fourth, and fifth modes at the zone center Γ for the lower frequency phonons (< 250 cm$^{-1}$), related to the optical phonon mode from breathing vibrations of oxygen coupled with the shocking modes of rhenium as shown in Figure. S3. In contrast, the contributions to $\lambda$ from the different phonon modes are comparable for the higher frequency phonons (500-250 cm$^{-}$ [1]). As shown in Figure. 4A, the magnitude of EPC constant $\lambda$ exhibits a fourfold enhancement from 0.2 at 0 GPa to about 0.8 at 30 GPa, where the rotation angle approaches 30° in the R-I phase. Such a large $\lambda$ and enhanced $N(E_F)$ are in favor of superconductivity.

Indeed, superconductivity is found theoretically throughout the R-I phase based on the Eliashberg equation using the typical Coulomb pseudopotential and the calculated logarithmic average frequency of the phonon spectrum $\omega_{\log}$ given in Figure. 4A. As shown in Figure. 4B, the $T_c$ starts to emerge above 10 GPa, increases quickly with pressure, and finally reaches ~15 K at 34 GPa where the octahedral rotation angle approaches 30° in the R-I phase. When we further increase the pressure to the phase boundary, where the rotation angle of the $ReO_6$ octahedra exceeds 30°, the R-I phase is a metastable state, as shown in Figure. 4A, B. Under these conditions, the evolutions of calculated $\lambda$ and $\omega_{\log}$ begin to change, and the calculated $T_c$ decreases above the critical pressure. The above results suggest that the densely packed oxygen lattice near the critical rotation angle is most favorable for superconductivity.

**Experimental verifications.** To verify the above theoretical predictions, we performed high-pressure resistance measurements on the $ReO_3$ single crystals by employing the cubic anvil cell (CAC) up to 13 GPa and diamond anvil cell (DAC) up to 65 GPa. As shown in Figure. S4A, B, the resistance and Hall resistance are consistent with previous reports at ambient pressure [21]. In Figure. S4C and Figure. 5A-C, we can see that the normal-state resistance shows a gradual increase with pressure in both C-II and R-I phases. Meanwhile, the resistance shows an obvious drop at low temperatures above 12 GPa, inset of Figure. S4C, which reaches zero resistance at 13 GPa, indicating the emergence of superconductivity in the R-I phase. The superconducting transition temperature first shows an enhancement in the R-I phase and then is gradually suppressed in the R-II phase. At 65.1 GPa, the superconducting transition cannot be distinguished down to 1.5 K.

The detailed evolution of superconducting transition can be clearly visualized from the normalized resistance $R(T)/R(20\ K)$ curves in Figure. 6A. Here, we define the temperatures of zero resistance as $T_c^{zero}$ and the deviation point from normal state resistance as $T_c^{onset}$. With increasing pressure gradually, the superconducting transition displays a continuous enhancement and the optimal $T_c^{onset}$ can reach ~ 17.1 K at 34 GPa. To the best of our knowledge, this is the highest $T_c$ among 5$d$ transition-metal oxides. Based on measurements in different runs, we constructed the superconducting $T_c(P)$ phase diagram, Figure. 6B, featured by a broad dome shape with an optimal $T_c \approx 17$ K at about 30 GPa. Combined with structural information, the optimal $T_c$ is achieved when the oxygen layers form the perfect hexagonal close packing. It is noted that the zero resistance was suppressed quickly in the R-II phase, and the observed resistance drops are likely caused by the superconducting R-I phase under over compression, consistent with the results of structural characterizations.

To further characterize the superconducting properties, we measured $R(T)$ under different magnetic fields at each pressure. As displayed in Figure. 6C and Figure. S7, the superconducting transition gradually shifts to lower temperatures with increasing magnetic field. Here, we use the 50% $R_n$ to define $T_c$. In Figure. 6D and Figure. S8, we can see the clearly upward curvature of the upper critical field near zero field, which indicates the multi-band feature of ReO$_3$ in the superconducting state. Therefore, we use the Werthamer-Helfand-Hohenberg (WHH) two-band model to fit $\mu_0 H_{c2}(T)$ [22]. The $\mu_0 H_{c2}(T)$ can be well fitted by the WHH two-band model, and the calculated $\mu_0 H_{c2}(0) \sim 14.6$ T is smaller than Pauli limit $\mu_0 H_p^{BCS} = 1.84 T_c = 28.3$ T, as shown in Figure. 6D. The above analysis indicates that the observed superconductivity in the R-I phase of ReO$_3$ should originate from multiple bands derived from the Re-5$d$ and O-2$p$ orbitals, consistent with the above DFT calculations.

It is noteworthy that the normal-state resistance of ReO$_3$ exhibits a continuous increase upon increasing pressures from R-I to R-II phase. To further track the correlation between structure phase transitions and transport properties, we plot the resistance values at 280 K and 20 K as a function of pressure in Figure. S5. As can be seen clearly, the resistance experiences obvious changes coinciding with the structure phase transitions and the emergence of superconductivity in the R-I phase. In addition, we performed Hall resistance measurements to obtain more information about the evolution of charge carriers in different phases of ReO$_3$ and to reveal its relationship with the emergence of superconductivity. The $R_H(H)$ curves measured at 20 K were displayed in Figure. 5D and the pressure dependence of carrier density $n$ extracted from $R_H = -1/(ne)$ was shown in Figure. 5E. According to previous studies, the C-I phase at AP contains both electron- and hole-carries with comparable density [21]. But the dominant charge carriers change to hole-type in the C-II and R-I phases, and its density jumps from C-II to R-I phase. Then, the dominant charge carriers change suddenly to electron-type again in the R-II phase. The Hall measurements suggest a significant reconstruction of the band structure accompanying the structure phase transitions, which can be reflected in the hole pocket around *L*-point in the calculated energy band at 30 GPa, as shown in Figure. S6. These changes should have a direct influence on the emergence of superconductivity in the R-I phase. These results indicate that the band structures undergo dramatic reconstruction accompanying the structural phase transitions in ReO$_3$, and the hole-type charge carriers dominate the superconducting state of the R-I phase.

We further analyze the normal-state transport properties of ReO$_3$ by using the modified Bloch-Grüneisen equation [15,23], which can provide more insights into the lattice dynamics. The $R(T)$ curves at both ambient and high pressures can be well fitted, and the obtained fitting results are displayed in Figure. 5F. As can be seen, both Debye and Einstein temperatures, $\Theta_D$ and $\Theta_E$, exhibit a dramatic drop when the structure transition takes place above ~ 10 GPa in the R-I phase. It is expected that such a phonon softening effect is beneficial for the emergence of superconductivity. This indicates that the low-frequency (< 500 cm$^{-1}$) phonon modes of oxygen in the R-I phase play a crucial role for the observed superconductivity, consistent with the above first-principles DFT calculations in Figure. 4A.

**DISCUSSION**

The recognition of closed-packed oxygen lattice in the HP R-I phase of $ReO_3$ enables us to discover superconductivity with a relatively high $T_c$ via combined theoretical and experimental investigations. The optimal $T_c$ of 17 K at ~ 30 GPa is higher than known rhenium-based oxides, including the superconducting $Cd_2Re_2O_7$ ($T_c \approx$ 1-2 K) [24] with the pyrochlore structure and $Hg_xReO_3$ ($T_c \approx$ 7.7 K @ 1atm and $\approx$ 11 K @ 4 GPa) [25] with the hexagonal bronze structure. To our knowledge, the optimal $T_c$ is also the highest among the existing 5$d$ transition-metal-based oxide superconductors. In sharp contrast to the aforementioned cases, the underlying mechanism for such a high $T_c$ should be largely attributed to the low-frequency phonon modes from the oxygen lattice. To quantify the contribution of light element to superconductivity, we made a rough comparison between the typical rare-earth superhydrides and $ReO_3$. According to Ref. 5, for instance, in $YH_9$ and $YH_{10}$, the contribution of hydrogen to $N(E_F)$ is approximately 24% and 28%, while its contribution to $\lambda$ is about 54% and 66%, respectively. In comparison, the corresponding contributions of oxygen to $N(E_F)$ and $\lambda$ are 27% and ~50% in the present case of $ReO_3$ at 30 GPa. In this regard, the R-I phase of $ReO_3$ can be considered as an oxide proxy to hydride superconductors that have aroused tremendous research interest recently.

**CONCLUSION**

In summary, the rhombohedral-I phase of $ReO_3$ was confirmed to be superconducting over a wide pressure range of 12-39 GPa, featured by a broad superconducting $T_c(P)$ dome with an optimal $T_c \approx 17$ K. In addition to the enhanced $N(E_F)$, the low-frequency phonon modes from the close-packed oxygen layers are found to play a crucial role in the observed superconductivity.

## MATERIALS AND METHODS

**Electrical transport measurements.** High-pressure resistance measurements of $ReO_3$ single crystals were carried out with a BeCu-type diamond anvil cell (DAC) and the soft KBr was employed as the pressure transmitting medium (PTM), and the $ReO_3$ single crystals were grown by the chemical vapor transport method [26]. Pressure values were determined at room temperature by the ruby fluorescence method [27] or from the first-order diamond Raman peak [28]. On the other hand, the temperature-dependent resistance measurements at low pressure range were carried out with a palm-type cubic anvil cell (CAC) up to 13 GPa and the glycerol was employed as the liquid PTM. The standard four probe method was used for resistance measurements. For Hall resistance measurements, the current was applied along <011> direction and the magnetic field was applied perpendicular to *ab* plane. The $R_H(H)$ data were anti-symmetrized with respect to the magnetic field between +4 and -4 T.

**X-ray diffraction measurements.** The high-pressure SXRD experiments were performed at the BL15U1 station of the Shanghai Synchrotron Radiation Facility (SSRF) ($\lambda = 0.6199$ Å). Here, the ethyl-alcohol mixture was used as a PTM with a ratio of 4:1. Using the geometric parameters calibrated by a $CeO_2$ standard, the DIOPTAS software package [29] was used to integrate the powder diffraction patterns and then convert them to one-dimensional profiles. Rietveld refinements of XRD patterns were conducted using GSAS with EXPGUI packages [30]. The *P-V* relationship is fitted by using the Birch-Murnaghan equation:

$$P(V) = \frac{3B_0}{2}\left[\left(\frac{V_0}{V}\right)^{7/3} - \left(\frac{V_0}{V}\right)^{5/3}\right]\left\{1 + \frac{3}{4}(B_0' - 4)\left[\left(\frac{V_0}{V}\right)^{2/3} - 1\right]\right\} \quad (1)$$

Where $V_0$ is the cell volume at ambient pressure, $V$ is the volume under pressure $P$, $B_0$ is the bulk elastic modulus, and $B_0' = 4$ is the derivative of the bulk elastic modulus at zero pressure. If we assume the rotation is rigid, the rotation angle $\theta$ can be estimated by the change of $a/c$, which can be obtained by geometric relation:

$$\theta = \arccos(\sqrt{6}\,a/c) \quad (2)$$

**Resistance fitting.** To probe the lattice dynamics, we analyzed the normal-state resistance by the following equation [15, 23], viz.

$$R(T) = R_0 + R_D\left(\frac{T}{\Theta_D}\right)^5 \int_0^{\Theta_D/T} \frac{z^5\, dz}{(e^z - 1)(1 - e^{-z})} + R_E\left[\frac{T}{\Theta_E}\sinh^2\left(\frac{\Theta_E}{2T}\right)\right]^{-1} \quad (3)$$

where the first term $R_0$ is the residual resistance, and the second and third terms are the Bloch-Grüneisen and Einstein expressions, which represent the electron scattering by acoustic and optical phonons, respectively. $R_D$ and $R_E$ are resistance constant and $\Theta_D$ and $\Theta_E$ are Debye and Einstein temperatures, respectively. The fitting results of CAC s1 are shown in Fig. S4C.

**WHH two-band model.** The Werthamer-Helfand-Hohenberg (WHH) two-band model in the dirty limit [22]:

$$a_0[\ln t + U(D_1 h)][\ln t + U(D_2 h)] + a_2[\ln t + U(D_2 h)] + a_1[\ln t + U(D_1 h)] = 0 \quad (4)$$

Where $t = T/T_c$ is the reduced temperature, $h = \hbar\mu_0 H_{c2}/2\phi_0 k_B T$ is the dimensionless magnetic field with $\phi_0$ the magnetic flux quantum, and $D_n$ is the charge carriers diffusion coefficient. The coefficients $a_0$-$a_2$ are defined as $a_0 = 2\omega/\lambda_0$, $a_1 = 1 + \lambda_-/\lambda_0$, $a_2 = 1 - \lambda_-/\lambda_0$ and $\lambda_- = \lambda_{11} - \lambda_{22}$, $\lambda_0 = \sqrt{\lambda_-^2 + 4\lambda_{12}\lambda_{21}}$, $\omega =$ $\lambda_{11}\lambda_{22} - \lambda_{12}\lambda_{21}$ where $\lambda$ is the superconducting coupling constants matrix. $U(x) = \psi(1/2 + x) - \psi(1/2)$, where $\psi(x)$ is the di-gamma function.

**DFT calculations.** The system was theoretically calculated by employing the first principles mothed as implemented in QUANTUM ESPRESSO (QE) package [31], and the equilibrium structure, phonon dispersions, EPC, and superconductivity of the compound were obtained as a function of the external hydrostatic pressure from 0 to 34 GPa. The plane-wave kinetic-energy cutoff and the energy cutoff for charge density were set as 100 and 400 Ry, respectively. The Brillouin zone (BZ) was sampled by a k-point mesh of $8 \times 8 \times 8$, and a Methfessel-Paxton smearing width of 0.02 Ry was adopted for calculating the self-consistent electron density. The dynamic and EPC matrix elements were computed with an $8 \times 8 \times 8$ $q$ mesh. The phonon and EPC were evaluated quantitatively by employing the density functional perturbation theory (DFPT) and Eliashberg theory [32,33].

The electronic structure, such as the density of state (DOS), was calculated using the Vienna Ab initio Simulation Package (VASP) [34], and the Perdew-Burke-Ernzerhof (PBE) flavor of the functional within the generalized gradient approximation (GGA) framework was employed to describe the electron-ion and electron-electron interactions [35, 36]. The cut-off energy of the plane wave was 520 eV, and the k-point density was at least 3600/ (number of atoms in the unit cell) in order to yield accurate results.

**Magnetic susceptibility measurement.** Magnetic susceptibility measurements were performed in an MPMS (Magnetic Property Measurement System) from Quantum Design. The miniature non-magnetic BeCu-type DAC with 400 μm culet was used, and the results were shown in Figure. S9.

# REFERENCES

[1] Drozdov, A.P., Kong, P.P., Minkov, V.S., et al. (2019). Superconductivity at 250 K in lanthanum hydride under high pressures. Nature *569*, 528-531. DOI:10.1038/s41586-019-1201-8.

[2] Somayazulu, M., Ahart, M., Mishra, A.K., et al. (2019). Evidence for superconductivity above 260 K in lanthanum superhydride at megabar pressures. Phys. Rev. Lett. *122*, 027001. DOI: 10.1103/PhysRevLett.122.027001.

[3] Drozdov, A.P., Eremets, M.I., Troyan, I.A., et al. (2015). Conventional superconductivity at 203 K at high pressures. Nature *525*, 73-76. DOI: 10.1038/nature14964.

[4] Bardeen, J., Cooper, L.N., and Schrieffer, J.R. (1957). Theory of superconductivity. Phys. Rev. *108*, 1175–1204. DOI: 10.1103/PhysRev.108.1175.

[5] Peng, F. Sun, Y., Pickard, C.J., et al. (2017). Hydrogen clathrate structures in rare earth hydrides at high pressures: possible route to room-temperature superconductivity. Phys. Rev. Lett. *119*, 107001. DOI: 10.1103/PhysRevLett.119.107001.

[6] Liu, H., Naumov, I.I., Hoffmann, R., et al. (2017). Potential high-$T_c$ superconducting lanthanum and yttrium hydrides at high pressure. Proc. Natl Acad. Sci. USA *114*, 6990-6995. DOI: 10.1073/pnas.1704505114.

[7] Duan, D.F., Liu, Y., Tian, F., et al. (2014). Pressure-induced metallization of dense $(H_2S)_2H_2$ with high-$T_c$ superconductivity. Sci. Rep. *4*, 6968. DOI: 10.1038/srep06968.

[8] Flores-Livas, J.A., Boeri, L., Sanna, A., et al. (2020). A perspective on conventional high-temperature superconductors at high pressure: methods and materials. Phys. Rep. *856*, 1–78. DOI:10.1016/j.physrep.2020.02.003.

[9] An, J.M., and Pickett, W.E. (2001). Superconductivity of $MgB_2$: Covalent Bonds Driven Metallic. Phys. Rev. Lett. *86*, 4366. DOI:10.1103/PhysRevLett.86.4366.

[10] Kortus, J., Mazin, I.I., Belashchenko, K.D., et al. (2001). Superconductivity of Metallic Boron in $MgB_2$. Phys. Rev. Lett. *86*, 4656. DOI: 10.1103/PhysRevLett.86.4656.

[11] Csányi, G., Littlewood, P.B., Nevidomskyy, A.H., et al. (2005). The role of the interlayer state in the electronic structure of superconducting graphite intercalated

compounds. *Nature Phys*. **1**, 42-45. DOI:10.1038/nphys119.

[12] Ferretti, A., Rogers, D.B., and Goodenough, J.B., (1965). The relation of the electrical conductivity in single crystals of rhenium trioxide to the conductivities of $Sr_2MgReO_6$ and $Na_xWO_3$. J. Phys. Chem. Solids. **26**, 2007. DOI:10.1016/0022-3697(65)90237-4.

[13] Evans, H.A., Wu, Y., Seshadri, R., and Cheetham, A.K. (2020). Perovskite-related $ReO_3$-type structures. Nat. Rev. Mater. **5**, 196. DOI:10.1038/s41578-019-0160-x.

[14] Falke, J. Chang, C.F., Liu, C.E., et al. (2021). Electronic structure of the metallic oxide $ReO_3$. Phys. Rev. B *103*, 115125. DOI:10.1103/PhysRevB.103.115125.

[15] Allen, P.B. and Schulz, W.W. (1993). Bloch-Boltzmann analysis of electrical transport in intermetallic compounds: $ReO_3$, $BaPbO_3$, $CoSi_2$, and $Pd_2Si$. Phys. Rev. B *47*, 14434. DOI:10.1103/PhysRevB.47.14434.

[16] Dyuzheva, T.I., Bendeliani, N.A., Glushko, A.N., et al. (1989). Phase diagram of $ReO_3$ up to 10 GPa. Phys. Scripta. *39*, 341. DOI:10.1088/0031-8949/39/3/013.

[17] Jorgensen, J.-E., Olsen, S.J., and Gerward, L. (2000). Phase transitions in $ReO_3$ studied by high-pressure X-ray diffraction. J. Appl. Crystallogr. *33*, 279. DOI:10.1107/S0021889899016659.

[18] Suzuki, E., Kobayashi, Y., Endo, S., et al. (2002). Structural phase transition in $ReO_3$ under high pressure. J. Phys. Condens. Matter. *14*, 10589. DOI:10.1088/0953-8984/14/44/338.

[19] Jorgensen J.-E., Marshall W.G., Smith R.I., et al. (2004). High-pressure neutron powder diffraction study of the *Im*-3 phase of $ReO_3$. J. Appl. Crystallogr. *37*, 857. DOI:10.1107/S0021889804018758.

[20] Muthu, D.V.S., Teredesai, P., Saha, S., et al. (2015). Pressure-induced structural phase transitions and phonon anomalies in $ReO_3$: Raman and first-principles study. Phys. Rev. B *91*, 224308. DOI:10.1103/PhysRevB.91.224308.

[21] Chen, Q., Lou, Z, Zhang, S.N., et al. (2021). Extremely large magnetoresistance in the "ordinary" metal $ReO_3$. Phys. Rev. B *104*, 115104. DOI:10.1103/PhysRevB.104.115104.

[22] Gurevich, A. (2003). Enhancement of the upper critical field by nonmagnetic impurities in dirty two-gap superconductors. Phys. Rev. B *67*, 184515. DOI:10.1103/PhysRevB.67.184515.

[23] Tanaka, T. Akahane, T., Bannai, E., et al. (1976). Role of polar optical phonon scattering in electrical resistivities of $LaB_6$ and $ReO_3$ (metallic conduction). J. Phys. C: Solid State Phys. *9*, 1235. DOI:10.1088/0022-3719/9/7/014.

[24] Hanawa, M. Muraoka, Y., Tayama, T., et al. (2001). Superconductivity at 1 K in

$Cd_2Re_2O_7$. Phys. Rev. Lett. *87*, 187001. DOI:10.1103/PhysRevLett.87.187001.

[25] Ohgushi, K. Yamamoto, A., Kiuchi, Y., et al. (2011). Superconducting phase at 7.7 K in the $Hg_xReO_3$ compound with a hexagonal bronze structure. Phys. Rev. Lett. *106*, 017001. DOI:10.1103/PhysRevLett.106.017001.

[26] Pearsall, T., (1973). On the growth of single crystals of $ReO_3$ by chemical vapour transport. J. Cryst. Growth *20*, 192. DOI: 10.1016/0022-0248(73)90003-1.

[27] Xu, J. A., Mao, H.K., and Bell, P.M. (1986). High-pressure ruby and diamond fluorescence: observations at 0.21 to 0.55 terapascal. Science *232*,1404-1406. DOI:10.1126/science.232.4756.14.

[28] Akahama, Y., and Kawamura, H. (2006). Pressure calibration of diamond anvil Raman gauge to 310 GPa. J. Appl. Phys. *100*, 043516. DOI:10.1063/1.2335683.

[29] Prescher, C., and Vitali, B.P. (2015). DIOPTAS: a program for reduction of two-dimensional X-ray diffraction data and data exploration, High Press. Res. *35*, 223-230. DOI:10.1080/08957959.2015.1059835.

[30] Toby, B.H. (2001). *EXPGUI*, a graphical user interface for *GSAS*. J. Appl. Crystallogr. *34*, 210. DOI:10.1107/S0021889801002242.

[31] Giannozzi, P. Baroni, S., Bonini, N., et al. (2009). QUANTUM ESPRESSO: a modular and open-source software project for quantum simulations of materials, J. Phys.: Condens. Matter *21* 395502. DOI:10.1088/0953-8984/21/39/395502.

[32] Baroni, S., de Gironcoli, S., Dal Corso, A., et al. (2001). Phonons and related crystal properties from density-functional perturbation theory, Rev. Mod. Phys. *73* 515. DOI:10.1103/RevModPhys.73.515.

[33] Giustino, F. (2017). Electron-phonon interactions from first principles, Rev. Mod. Phys. *89* 015003. DOI:10.1103/RevModPhys.89.015003.

[34] Kresse, G. and Furthmüller, J. (1996). Efficient iterative schemes for ab initio total-energy calculations using a plane-wave basis set, Phys. Rev. B *54* 11169. DOI:10.1103/PhysRevB.54.11169.

[35] Perdew, J.P., Burke, K., and Ernzerhof, M. (1996). Generalized gradient approximation made simple, Phys. Rev. Lett. *77* 3865. DOI:10.1103/PhysRevLett.77.3865.

[36] Blöchl, P.E. (1994). Projector augmented-wave method, Phys. Rev. B, *50* 17953. DOI:10.1103/PhysRevB.50.17953.

**ACKNOWLEDGMENTS**

This work is supported by National Key R&D Program of China (2023YFA1406100, 2022YFA1403900 and 2021YFA1400200), National Natural Science Foundation of China (12025408, 11921004, 12174424, 92065201, 12304030, U23A6003), the

Strategic Priority Research Program of CAS (XDB33000000), the CAS Project for Young Scientists in Basic Research (2022YSBR-047 and 2022YSBR-048), the CAS PIFI program (2024PG003), the Youth Innovation Promotion Association of CAS (2023007 and 2018010), and the Project of China Postdoctoral Science Foundation (Grant No. 2023M743223). We thank the staff members of the BL15U1 station (https://cstr.cn/31124.02.SSRF.BL15U1) and User Experiment Assist System in Shanghai Synchrotron Radiation Facility (SSRF) (https://cstr.cn/31124.02.SSRF) for providing technical support and assistance in data collection and analysis. Part of the high-pressure resistance measurements were performed at the Cubic Anvil Cell station (https://cstr.cn/461 31123.02.SECUF.A2) of Synergic Extreme Condition User Facility (SECUF) (https://cstr.cn/31123.02.SECUF).

## AUTHOR CONTRIBUTIONS

J.C. designed research. P.S. performed high pressure XRD. P.S., Z.L., P.Y. and J.H. performed high pressure transport measurements. Y.J. grew crystals. T.L., L.M. and M.L. carried out DFT calculations. All authors participated in analyzing the data. P.S. and J.C. wrote the original manuscript. J.C. and J.S. reviewed and edited articles. J.C. supervised the research work. All the authors reviewed and edited the manuscript.

## DECLARATION OF INTERESTS

The authors declare no competing interests.

# Figures and captions

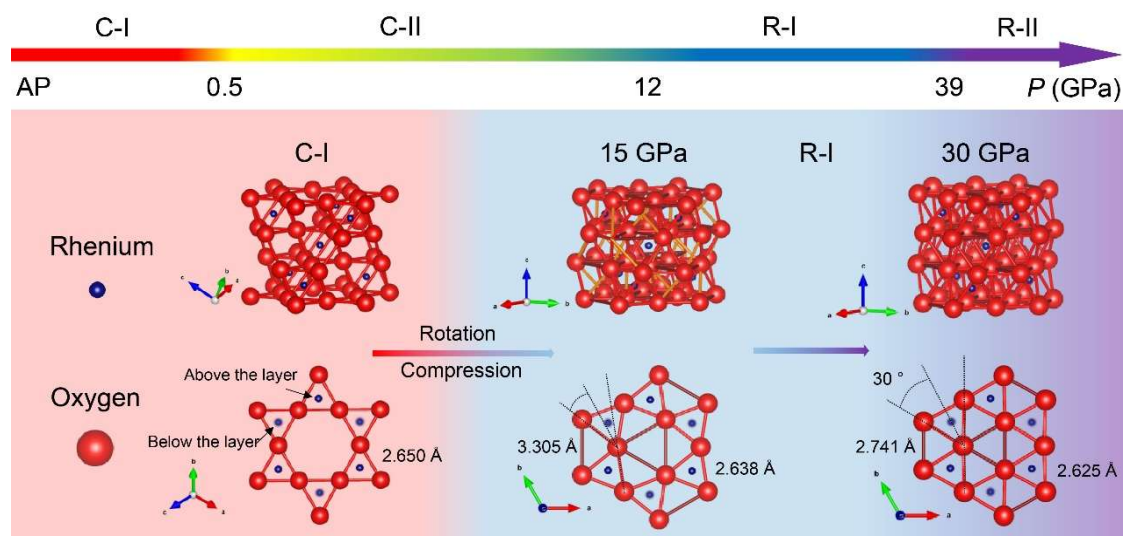

**Figure. 1 Crystal structure evolution of ReO$_3$ under high pressure.** The color bar indicates the critical pressures separating different structural phases. Re cations are shown in blue, and O anions in red. The top panel illustrates the crystal structures of ReO$_3$ from the C-I to R-I phases at 0 GPa, 15 GPa, and 30 GPa. The bottom panel depicts the coupled rotation of oxygen layers in the (111) plane, transitioning from a Kagome lattice to a triangular lattice. The faded Re cations lie below the oxygen layer, while the bright Re cations are positioned above the oxygen layer. Bond lengths were obtained from DFT calculations, as detailed in Table S2.

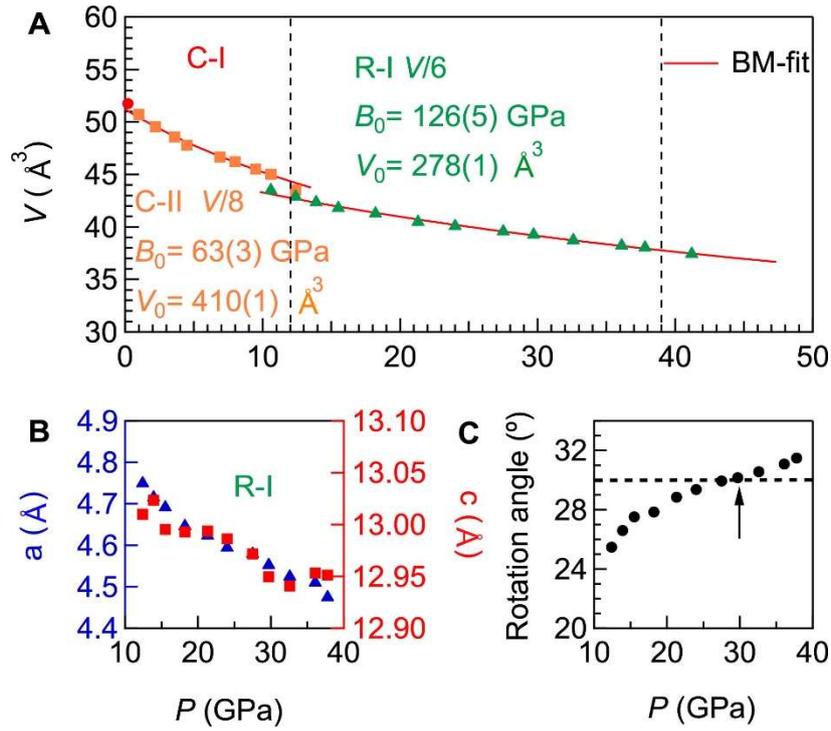

**Figure. 2 Refined lattice parameters of ReO₃ under pressure.** (A) Pressure dependence of volume in the range of 0.2–41.2 GPa for different structural phases of ReO$_3$. Solid red lines represent the Birch-Murnaghan (BM) equation of state (EoS) fitting, while the dashed line marks the phase boundaries. (B) Refined lattice parameters as a function of pressure in the R-I phase. Blue triangles denote the *a*-axis, and red squares denote the *c*-axis. (C) Pressure-dependent coupled rotation angle of ReO$_6$ octahedra in the R-I phase. The dashed line and black arrow indicate the critical pressure at which the oxygen layers form a perfect hexagonal-close-packed lattice.

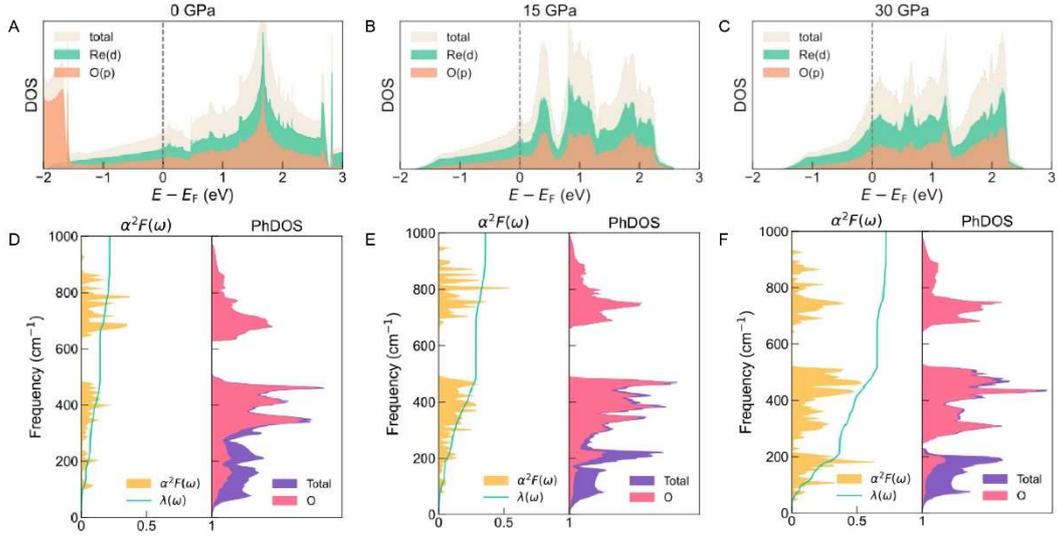

**Figure. 3 Calculated electronic density of states and EPC properties.** (A-C) Total density of states (DOS) and partial DOS for ReO$_3$ at 0, 15, and 30 GPa. The energy zero represents the Fermi level. (D-F) Eliashberg spectral function $α^2F(ω)$, and cumulative frequency-dependent electron-phonon coupling (EPC) constant $λ(ω)$, total phononic density of states (PhDOS) and partial PhDOS at 0, 15, and 30 GPa.

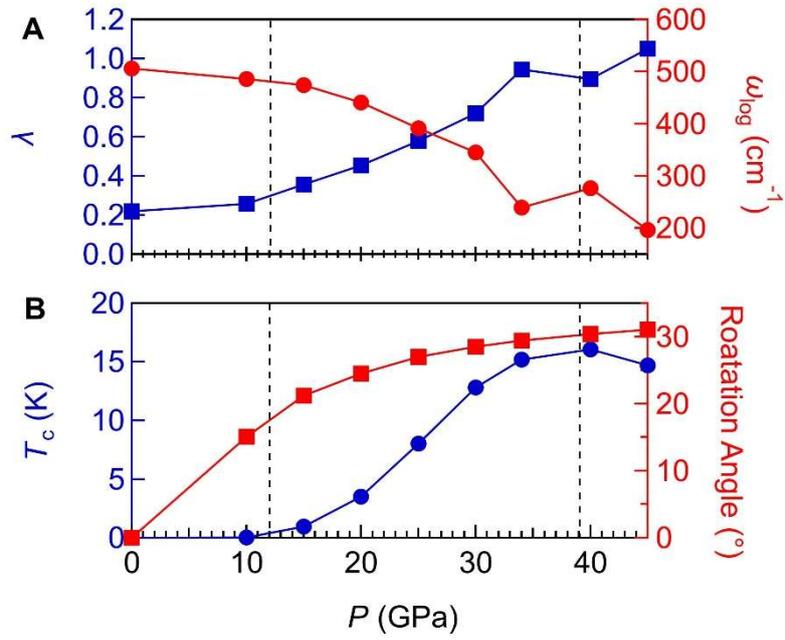

**Figure. 4 Pressure effects on superconducting properties.** (A) Calculated $\lambda$, $\omega_{\log}$, (B) $T_c$, and rotation angle as a function of pressure. The dashed line marks the phase boundaries.

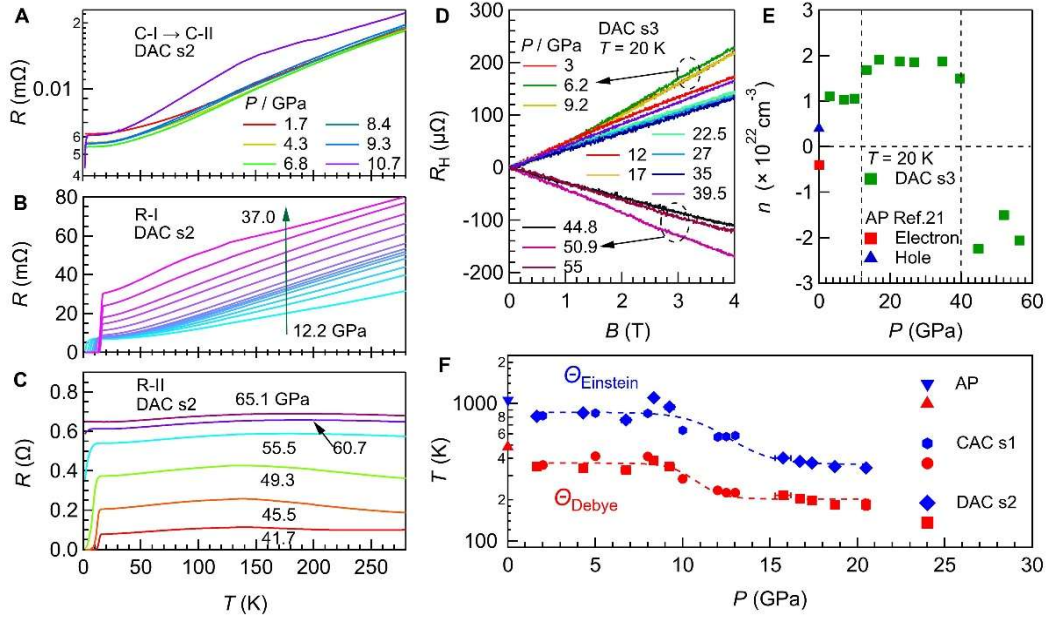

**Figure. 5 Electrical transport properties under pressure.** Temperature dependence of resistance of ReO$_3$ (DAC s2) measured at various pressures (A) from 1.7 to 10. 7 GPa, (B) from 12.0 to 37 GPa, and (C) from 41.7 GPa to 65.1 GPa. (D) Field dependence of Hall resistance (DAC s3) measured at 20 K under various pressures up to 55 GPa. (E) Pressure dependence of the carrier concentration at 20 K. The data at ambient pressure is adopted from Ref. [21]. The dashed line marks the phase boundaries. (F) Pressure dependence of the estimated Debye and Einstein temperatures from fitting to the modified Bloch-Grüneisen equation.

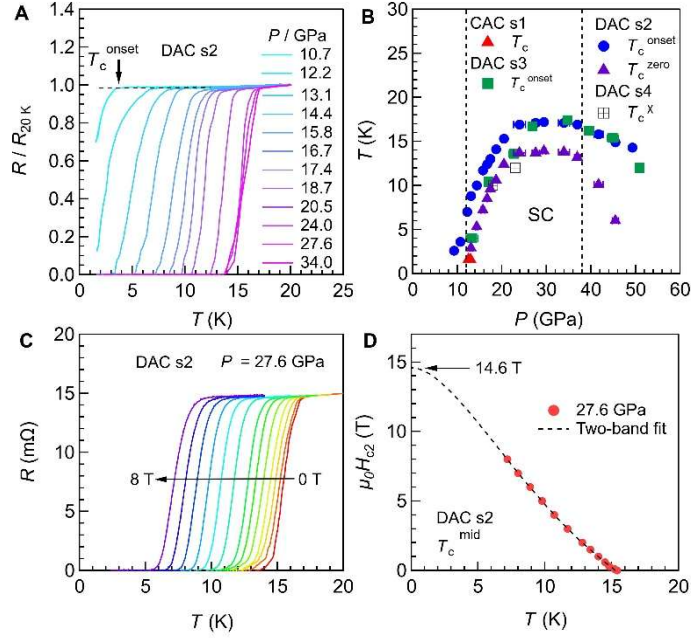

**Figure. 6. The superconductivity property.** (A) Normalized resistance $R/R_{20\ K}$ of ReO$_3$ (DAC s2) under various pressures up to 34 GPa. (B) Temperature-pressure phase diagram of ReO$_3$ in different runs. The dashed line marks the phase boundaries. (C) Temperature dependence of resistance under various magnetic fields at 27.6 GPa. (D) Temperature dependence of the upper critical field $\mu_0 H_{c2}(T)$ at 27.6 GPa. The dashed line represents the fitting results by using WHH two-band model.